\documentclass[prc,letterpaper,showpacs,preprint]{revtex4}

\usepackage{graphicx}
\usepackage{amsmath}
\usepackage{bm}

\begin{document}

\preprint{ \vbox{\hbox{ JLAB-THY-06-493}}}

\title{Conserved electromagnetic currents in a relativistic optical model}

\author{ J.~W.~Van~Orden}

\affiliation{Department of Physics, Old Dominion University,
Norfolk, VA 23529\\and\\Jefferson Lab\footnote{Notice:  This
manuscript has been authored by The Southeastern Universities
Research Association, Inc. under Contract No. DE-AC05-84ER40150 with
the U. S. Department of Energy.  The United States Government
retains and the publisher, by accepting the article for publication,
acknowledges that the United States Government retains a
non-exclusive, paid-up, irrevocable, world wide license to publish
or reproduce the published form of this manuscript, or allow others
to do so, for United States Government purposes.}, 12000 Jefferson
Avenue, Newport News, VA 23606
 }

\date{\today}

\begin{abstract}
A simple model of a relativistic optical model is constructed by
reducing the three-body Bethe-Salpeter equation to an effective
two-body optical model. A corresponding effective current is derived
for use with the optical-model wave functions. It is shown that this
current satisfies a Ward-Takahashi identity involving the optical
potential which results in conserved current matrix elements.
\end{abstract}

\pacs{25.30.Fj, 24.10.Jv, 24.10.Ht, 21.45.+v, 24.10.Cn}

\maketitle

\section{Introduction}

Dirac optical models are widely used in analyzing electron
scattering data for $(e,e')$ and $(e,e'p)$ reactions\cite{PVOW85,
PVO87,PVO89,CPVO89a,CPVO89b,CP92,JOW92,JO,KW97,KW99,KW03,USMGC93,
USMGC95,USMGC96,UCMAD99,UV00,UCMVE01,MCDMU04,VMCMU04,MLJRVU06,MGP01a,MGP01b,
MGP02,MCGP03}. Recently these models have shown to give excellent
agreement with spin observables for $^{16}O(e,e'p)$ \cite{Fissum}
where the behavior of the observables at large missing momentum has
been attributed to dynamical relativistic effects due to ``spinor
distortion''. These models have also been used in the analysis of
recent data for $^4He(e,e'p)$ \cite{Dieterich,Strauch,Aniol} for
indications of medium modification of nucleons in the nuclei.
Evidence for such modifications in this case relies on use of
optical model calculations with and without medium modified from
factors. Since the size of the difference between the calculation
without medium modified form factors and the data is on the order of
5 to 10 percent, any conclusion based on this approach requires that
the model calculations can be trusted to a similar level of
accuracy. (It should be noted that this later effect has been
described in a more traditional approach \cite{SBKMV05} by including
charge exchange interactions. The choice of the parameters for are
reasonable, but are not well constrained by data.)

Given the importance of the questions that are being addressed with
these models, it is necessary to consider their foundations. The
fundamental assumption is that the nucleon-nucleus interaction can
be described in terms of a single-particle hamiltonian of the form
\begin{equation}
H=\frac{1}{i}\bm{\alpha}\cdot\bm{\nabla}+\beta\left[m
+S_{OPT}(r,E)\right]+V_{OPT}(r,E)\,,\label{DiracOPT}
\end{equation}
where $S_{OPT}(r,E)$ and $V_{OPT}(r,E)$ are complex,
energy-dependent, scalar and vector optical potentials. This
approach was first used to provide a phenomenological description of
proton-nucleus elastic scattering\cite{AC79,ACM79,ACMS81}. It was
subsequently demonstrated that optical potentials derived from
parameterized NN interactions in the impulse approximation provided
a very good description of the spin observables for proton-nucleus
elastic scattering at intermediate energies with a minimal number of
parameters\cite{SMW93,CHMRS,HPTT84}. It should be noted, however,
that although the origins of the optical potential in
nonrelativistic multiple-scattering theory have received a great
deal of theoretical attention, the Dirac optical model proceeded by
analogy to the nonrelativistic case without reference to a
relativistic many-body theory.

Similarly, the first applications of the Dirac optical model to
$(e,e'p)$ and $(e,e')$ reactions
\cite{PVOW85,PVO87,PVO89,CPVO89a,CPVO89b} assumed that the necessary
current matrix elements could be obtained in analogy to the
nonrelativistic case with wave functions obtained from one-body
Dirac equations and the current operator described by a one-body
current. In both the nonrelativistic and relativistic cases this
assumption leads to a lack of current conservation.  This lack of
current conservation is a direct result of the underlying many-body
nature of these reactions. This manifests itself in several related
ways. The first is associated with the composite nature of the
nucleon resulting in the need for form factors which interfere with
the usual single-particle Ward-Takahashi identities and imply that
one-body current be of a much more complicated general off-shell
form. The second is associated the with the appearance of many-body
exchange or interaction currents. Finally, the use of an optical
potential implies that the many-body problem has been reduced to an
effective two-body where the contributions of channels associated
with excitation of the residual system are hidden in the optical
potentials. Since these hidden channels can be excited by virtual
photon absorption, a consistent treatment of the reaction requires
that an effective current operator be used in place of the simple
one-body current. The first source of current non-conservation has
been addressed by studying the effect of various on-shell equivalent
forms of the single-nucleon current on the optical model
calculations\cite{PVOW85,CP92,USMGC93,UV00}. This gives some rough
indication of the size of violation of current conservation, but
does not really address the underlying problem. The second source
has been addressed by including two-body meson-exchange currents in
an approximate fashion\cite{MGP02}. The problem of the effect of the
reduction of the many-body problem to an effective optical model on
the current has been discussed in a general fashion, but has not
been studied in any concrete way.\cite{CPVO89a,BVO90}

The purpose of this paper is to show that it is indeed possible to
obtain a Dirac optical model from an underlying covariant theory and
to obtain the corresponding effective current operator necessary to
maintain electromagnetic current conservation. In doing this several
choices will be made in the reorganization of the covariant theory
into the optical model.  Clearly, this approach is not necessarily
unique. Therefore, the hope is that this work will stimulate the
development of alternate approaches with the hope that this will
lead to an improvement in the phenomenology for the application of
the Dirac optical model to electromagnetic processes.

The starting point for this work are the many-body Bethe-Salpeter
equations. These equations are most easily understood as a
resummation of all Feynman diagrams for n-point functions. The $n/2$
particles associated with the external legs of the n-point function
are treated as explicit degrees of freedom while all other degrees
of freedom are collected into a set of irreducible kernels. These
degrees of freedom are implicit. The kernels are then used in
integral equations to sum all contributions to the n-point
functions. Since these equations are based in Feynman perturbation
theory, all elements of the integral equations are manifestly
covariant. For spin-1/2 constituents the one-body propagators
appearing in the integral equations are solutions to the Dirac
equation so it reasonable to believe that it is be possible to
reduce the many-body problem to an effective theory involving the
interaction of a Dirac particle with an $(n-1)$-body system. The
structure of the integral equation for the Bethe-Salpeter $n$-point
functions is similar in form to those for nonrelativistic
multiple-scattering theory with the exceptions that all integrals
are four-dimensional rather than three dimensional, and that the all
propagators are local, whereas propagators of time-ordered
description usually used in multiple-scattering theory are global.

For simplicity, the simplest illustrative case of the process, the
three-body Bethe-Salpeter equation\cite{Taylor66,norm} for
distinguishable particles, is used to show how the reduction to an
effective optical model can be implemented. This equation is
relatively simple in structure and the construction of
electromagnetic current matrix elements for this equation is well
understood\cite{kb97II,kb99,3NCur}. The optical model is obtained by
reducing the three-body problem to an effective two-body problem.
The effective kernel for the interaction between the bound state of
two of the particles and the remaining particle can then be
interpreted as an ``optical potential.'' A similar reduction of the
Bethe-Salpeter current matrix elements leads to the identification
of an effective current operator consistent with the optical model.
This optical model current will be shown to result in conserved
current matrix elements.

In the first section  the optical model for the interaction of one
particle with a bound state of the remaining pair is constructed.
Next, bound and scattering states are defined in terms of the
optical model states. Finally, the effective optical model current
is constructed and the impulse approximation contribution to the
effective current is isolated. It is then shown that the optical
model current satisfies a Ward-Takahashi identity involving the
optical potential which results in conserved current matrix
elements.

\section{Optical Model Representation of the Three-body Scattering
Matrix}

Here we will use a matrix form for the three-body Bethe-Salpeter
equation described in \cite{3NCur} to simplify the reduction of the
three-body problem to the effective two-body problem. This is
formulation is summarized in the appendix for the convenience of the
reader. For three distinguishable particles, the three-body
scattering matrix can be written in matrix form as using
(\ref{t_BS})
\begin{equation}
\bm{{\cal T}}= \bm{{\cal M}}-\bm{{\cal
M}}\bm{G}^0_{BS}\bm{{B}}\bm{{\cal T}}\,,\label{t_BS_0}
\end{equation}
where the matrices are defined in the appendix.

Our objective is to reduce this expression so than we can extract an
effective equation for particle 1 scattering from a bound state of
particles 2 and 3. This is accomplished by separating the two-body t
matrix for particles 2 and 3 into terms containing bound-state poles
and a residual piece containing only the scattering cuts. Assuming
that there is only a single bound state for particles 2 and 3, the
two-body scattering matrix in momentum space has the form
\begin{eqnarray}
M^1(p^{1\prime},p^1,P^1)&=&\frac{1}{2E(\bm{P})}\left(\frac{\Gamma^{(2)1}(p^{1\prime},\hat{P}^1)
{\Gamma^{(2)1}}^\dag(p^1,\hat{P}^1)}{{P^1}^0-E(\bm{P}^1)+i\eta}-\frac{\Gamma^{(2)1}(-p^{1\prime},-\hat{P}^1)
{\Gamma^{(2)1}}^\dag(-p^1,-\hat{P}^1)}{{P^1}^0+E(\bm{P}^1)-i\eta}\right)\nonumber\\
&&+ M^1_r(p^{1\prime},p^1,P^1)\,,\label{Mdecomp0}
\end{eqnarray}
where $P^1$ is the total four-momentum of the pair, $p^1$ and
$p^{1\prime}$ are the initial and final relative four-momenta of the
pair, $\Gamma^{(2)1}$ is the bound-state vertex function for the
pair, $m^1$ is the mass of the bound state and $M^1_r$ is the
residual scattering matrix. For relativistic many-body equations
there is no clean factorization of the vertex functions into
relative and center-of-mass pieces. This means that vertex functions
are explicitly dependent upon the total momentum. Any spinor indices
associate with the vertex function are suppressed and are assumed to
be summed. Note that there are positive and negative poles
associated with the positive- and negative-energy bound states.
There are several ways to precede at this point in reducing the
three-body problem to an effective two-body problem. Both the
positive and negative poles can be retained and thus explicitly
include the interaction with particle 1 and the negative-energy
bound state. This has the virtue that the decomposition of the
scattering matrix can be written in a manifestly covariant form.
However, any attempt to write this in the form of an optical model
will result in a form which is more complicated than is usually
assumed. In addition, it is reasonable to assume that the
contributions from the negative-energy pole will be small especially
when this approach is extended to systems with more particles which
means that the additional complexity may have little real physical
impact.

A second approach would then be to treat only the positive-energy
pole with the negative energy pole becoming part of the residual
scattering matrix. This decomposition will not be manifestly
covariant. Furthermore, the vertex functions are only uniquely
defined at the pole and (\ref{Mdecomp0}) assumes that they are
defined at this point as is indicated by the hat over the total
momentum. The momenta in (\ref{t_BS_0}) are not similarly restricted
and the action of inverse two-body propagators on the bound state
are not generally defined.

A third alternative, which is the one used here, is to assume that
in any loop integrals involving the bound state the positive energy
pole will be picked up resulting which will restrict $P^1$ to be
on-shell. This prescription is manifestly covariant and the
resulting decomposition of the two-body scattering matrix is also
covariant. This can be realized in the equations for the three-body
scattering matrix by writing
\begin{equation}
\bm{{\cal M}}=-\bm{D}^1\left|\Gamma^{(2)1}\right>iG_1^{-1}(i{\cal
Q}^1) \left<\Gamma^{(2)1}\right|{\bm{D}^1}^T+\bm{{\cal
M}}_R=\bm{{\cal M}}_P+\bm{{\cal M}}_R\,,\label{Mdecomp}
\end{equation}
where
\begin{equation}
\bm{D}^{1}=\left(
\begin{array}{c}
0 \\
 1 \\
 0\\
  0
\end{array}
\right)
\end{equation}
and ${\cal Q}^1$ is an operator that places the total momentum for
particles 2 and 3 on the bound-state mass shell by requiring that
the appropriate pole be picked up in any loops containing the
two-body t-matrix for particles 2 and 3.

We can now decompose the equation for the t-matrix according to
whether the initial  (final) interaction is the pole contribution
(superscript $P$) or the residual contribution (superscript $R$.)
This leads to the set of coupled matrix equations
\begin{eqnarray}
\bm{{\cal T}}^{PP}&=&\bm{{\cal M}}_P-\bm{{\cal
M}}_P\bm{G}^0_{BS}\bm{{B}}
\left(\bm{{\cal T}}^{PP}+\bm{{\cal T}}^{RP}\right)\label{TPP}\\
\bm{{\cal T}}^{RP}&=&-\bm{{\cal M}}_R\bm{G}^0_{BS}\bm{{B}}
\left(\bm{{\cal T}}^{PP}+\bm{{\cal T}}^{RP}\right)\label{TRP}\\
\bm{{\cal T}}^{PR}&=&-\bm{{\cal M}}_P\bm{G}^0_{BS}\bm{{B}}
\left(\bm{{\cal T}}^{PR}+\bm{{\cal T}}^{RR}\right)\label{TPR}\\
\bm{{\cal T}}^{RR}&=&\bm{{\cal M}}_R-\bm{{\cal
M}}_R\bm{G}^0_{BS}\bm{{B}} \left(\bm{{\cal T}}^{PR}+\bm{{\cal
T}}^{RR}\right)\label{TRR}\,.
\end{eqnarray}
The solution of this set of equations is facilitated by defining a
scattering matrix that does not contain any contributions from the
bound state for particles 2 and 3. This is defined as
\begin{equation}
\bm{T}_R=\bm{{\cal M}}_R-\bm{{\cal
M}}_R\bm{G}^0_{BS}\bm{{B}}\bm{T}_R =\bm{{\cal
M}}_R-\bm{T}_R\bm{G}^0_{BS}\bm{{B}}\bm{{\cal M}}_R\,.
\end{equation}
The second of these two forms can be solved to give
\begin{equation}
\bm{{\cal M}}_R=\bm{T}_R+\bm{T}_R\bm{G}^0_{BS}\bm{{B}}\bm{{\cal
M}}_R\,.\label{MsubR}
\end{equation}
Using this, (\ref{TRP}) can be rewritten as
\begin{eqnarray}
\bm{{\cal
T}}^{RP}&=&-\left(\bm{T}_R+\bm{T}_R\bm{G}^0_{BS}\bm{{B}}\bm{{\cal
M}}_R\right)\bm{G}^0_{BS}\bm{{B}} \left(\bm{{\cal T}}^{PP}+\bm{{\cal
T}}^{RP}\right)\nonumber\\
&=&-\bm{T}_R\bm{G}^0_{BS}\bm{{B}} \left(\bm{{\cal T}}^{PP}+\bm{{\cal
T}}^{RP}\right)-\bm{T}_R\bm{G}^0_{BS}\bm{{B}}\bm{{\cal
M}}_R\bm{G}^0_{BS}\bm{{B}} \left(\bm{{\cal T}}^{PP}+\bm{{\cal
T}}^{RP}\right)\nonumber\\
&=&-\bm{T}_R\bm{G}^0_{BS}\bm{{B}} \left(\bm{{\cal T}}^{PP}+\bm{{\cal
T}}^{RP}\right)+\bm{T}_R\bm{G}^0_{BS}\bm{{B}}\bm{{\cal T}}^{RP}\nonumber\\
&=&-\bm{T}_R\bm{G}^0_{BS}\bm{{B}}\bm{{\cal T}}^{PP}\,,\label{TRP2}
\end{eqnarray}
where we have used the fact that ${\bm{D}^1}^T\bm{B}\bm{D}^1=0$ in
simplifying the equations. This can be used in (\ref{TPP}) to give
\begin{equation}
\bm{{\cal T}}^{PP}=\bm{{\cal M}}_P-\bm{{\cal
M}}_P\bm{G}^0_{BS}\bm{{B}}\bm{{\cal T}}^{RP}=\bm{{\cal
M}}_P+\bm{{\cal
M}}_P\bm{G}^0_{BS}\bm{{B}}\bm{T}_R\bm{G}^0_{BS}\bm{{B}}\bm{{\cal
T}}^{PP}\,.\label{TPP2}
\end{equation}
From (\ref{Mdecomp}),
\begin{equation}
\bm{{\cal M}}_P=-\bm{D}^1\left|\Gamma^{(2)1}\right>iG_1^{-1}(i{\cal
Q}^1)
\left<\Gamma^{(2)1}\right|{\bm{D}^1}^T=-iG_1^{-1}\bm{D}^1\left|\Gamma^{(2)1}\right>(-iG_1)i{\cal
Q}^1 \left<\Gamma^{(2)1}\right|{\bm{D}^1}^TiG_1^{-1}\,.
\end{equation}
Using this and iterating of (\ref{TPP2}) it is possible to make the
conjecture that
\begin{equation}
\bm{{\cal
T}}^{PP}=-iG_1^{-1}\bm{D}^1\left|\Gamma^{(2)1}\right>i{\cal
Q}^1(-iG_{OPT})i{\cal Q}^1
\left<\Gamma^{(2)1}\right|{\bm{D}^1}^TiG_1^{-1}\,.
\end{equation}
Substituting this into (\ref{TPP2}),
\begin{eqnarray}
&&-iG_1^{-1}\bm{D}^1\left|\Gamma^{(2)1}\right>i{\cal
Q}^1(-iG_{OPT})i{\cal Q}^1
\left<\Gamma^{(2)1}\right|{\bm{D}^1}^TiG_1^{-1}\nonumber\\
&&\qquad=-iG_1^{-1}\bm{D}^1\left|\Gamma^{(2)1}\right>(-iG_1)i{\cal
Q}^1 \left<\Gamma^{(2)1}\right|{\bm{D}^1}^TiG_1^{-1}\nonumber\\
&&\qquad+iG_1^{-1}\bm{D}^1\left|\Gamma^{(2)1}\right>(-iG_1)i{\cal
Q}^1
\left<\Gamma^{(2)1}\right|{\bm{D}^1}^TiG_1^{-1}\bm{G}^0_{BS}\bm{{B}}\bm{T}_R\bm{G}^0_{BS}
\bm{{B}}iG_1^{-1}\bm{D}^1\left|\Gamma^{(2)1}\right>i{\cal Q}^1
\nonumber\\
&&\qquad\times(-iG_{OPT})i{\cal
Q}^1\left<\Gamma^{(2)1}\right|{\bm{D}^1}^TiG_1^{-1}\,.
\end{eqnarray}
We can then identify
\begin{eqnarray}
i{\cal Q}^1G_{OPT}i{\cal Q}^1&=&G_1i{\cal Q}^1\nonumber\\
&&-G_1i{\cal Q}^1
\left<\Gamma^{(2)1}\right|{\bm{D}^1}^TiG_1^{-1}\bm{G}^0_{BS}\bm{{B}}\bm{T}_R\bm{G}^0_{BS}
\bm{{B}}iG_1^{-1}\bm{D}^1\left|\Gamma^{(2)1}\right>i{\cal
Q}^1G_{OPT}i{\cal Q}^1\nonumber\\
&=&G_1i{\cal Q}^1-G_1i{\cal Q}^1
\left<\Gamma^{(2)1}\right|{\bm{D}^1}^TG^{1}_{BS}\bm{{B}}\bm{T}_RG^{1}_{BS}
\bm{{B}}\bm{D}^1\left|\Gamma^{(2)1}\right>{\cal Q}^1G_{OPT}i{\cal
Q}^1
\end{eqnarray}
where $G^{1}_{BS}=-iG_2G_3$ is the free two-body propagator for
particles 2 and 3, and
$\left|\Phi^{(2)1}\right>=G^{1}_{BS}\left|\Gamma^{(2)1}\right>$ is
the bound-state Bethe-Salpeter wave function for particles 2 and 3.
Defining
\begin{equation}
V_{OPT}=i{\cal
Q}^1\left<\Phi^{(2)1}\right|{\bm{D}^1}^T\bm{{B}}\bm{T}_R
\bm{{B}}\bm{D}^1\left|\Phi^{(2)1}\right>{\cal Q}^1\label{defVopt}
\end{equation}
as the optical potential, the optical model propagator can be
rewritten as
\begin{equation}
i{\cal Q}^1G_{OPT}i{\cal Q}^1=\left(G_1-G_1
V_{OPT}G_{OPT}\right)i{\cal Q}^1\,.
\end{equation}
Note that keeping only the leading terms in \ref{defVopt} yields
\begin{equation}
V_{OPT}=i{\cal Q}^1\left<\Phi^{(2)1}\right|\left({\cal M}^0+{\cal
M}^2+{\cal M}^3\right)\left|\Phi^{(2)1}\right>{\cal Q}^1\,.
\end{equation}
With the exception of the explicit three-body term ${\cal M}^0$,
this is the impulse approximation to the optical potential.

Substituting (\ref{MsubR}) into (\ref{TRR}) gives
\begin{eqnarray}
\bm{{\cal T}}^{RR}&=&\bm{T}_R+\bm{T}_R\bm{G}^0_{BS}\bm{{B}}\bm{{\cal
M}}_R-\left(\bm{T}_R+\bm{T}_R\bm{G}^0_{BS}\bm{{B}}\bm{{\cal
M}}_R\right)\bm{G}^0_{BS}\bm{{B}} \left(\bm{{\cal T}}^{PR}+\bm{{\cal
T}}^{RR}\right)\nonumber\\
&=&\bm{T}_R+\bm{T}_R\bm{G}^0_{BS}\bm{{B}}\bm{{\cal
M}}_R-\bm{T}_R\bm{G}^0_{BS}\bm{{B}} \left(\bm{{\cal
T}}^{PR}+\bm{{\cal
T}}^{RR}\right)\nonumber\\
&&\qquad-\bm{T}_R\bm{G}^0_{BS}\bm{{B}}\bm{{\cal
M}}_R\bm{G}^0_{BS}\bm{{B}} \left(\bm{{\cal T}}^{PR}+\bm{{\cal
T}}^{RR}\right)\nonumber\\
&=&\bm{T}_R+\bm{T}_R\bm{G}^0_{BS}\bm{{B}}\bm{{\cal
M}}_R-\bm{T}_R\bm{G}^0_{BS}\bm{{B}} \left(\bm{{\cal
T}}^{PR}+\bm{{\cal
T}}^{RR}\right)+\bm{T}_R\bm{G}^0_{BS}\bm{{B}}\left(\bm{{\cal
T}}^{RR}-\bm{{\cal M}}_R\right)\nonumber\\
&=&\bm{T}_R-\bm{T}_R\bm{G}^0_{BS}\bm{{B}} \bm{{\cal
T}}^{PR}\,.\label{TRR1}
\end{eqnarray}
This can be used in (\ref{TPR}) to yield
\begin{eqnarray}
\bm{{\cal T}}^{PR}&=&-\bm{{\cal M}}_P\bm{G}^0_{BS}\bm{{B}}
\bm{{\cal T}}^{RR}\nonumber\\
&=&-\bm{{\cal M}}_P\bm{G}^0_{BS}\bm{{B}}
\left(\bm{T}_R-\bm{T}_R\bm{G}^0_{BS}\bm{{B}} \bm{{\cal T}}^{PR}\right)\nonumber\\
&=&-\bm{{\cal M}}_P\bm{G}^0_{BS}\bm{{B}}\bm{T}_R + \bm{{\cal
M}}_P\bm{G}^0_{BS}\bm{{B}}\bm{T}_R\bm{G}^0_{BS}\bm{{B}} \bm{{\cal
T}}^{PR}
\end{eqnarray}
Iteration of this shows that that it can be rewritten as
\begin{equation}
\bm{{\cal T}}^{PR}=-\bm{{\cal
T}}^{PP}\bm{G}^0_{BS}\bm{{B}}\bm{T}_R\,.\label{TPR2}
\end{equation}
Using (\ref{TPR2}) in (\ref{TRR1}),
\begin{equation}
\bm{{\cal T}}^{RR}=\bm{T}_R+\bm{T}_R\bm{G}^0_{BS}\bm{{B}} \bm{{\cal
T}}^{PP}\bm{G}^0_{BS}\bm{{B}}\bm{T}_R\,.\label{TRR2}
\end{equation}

The complete t-matrix is the sum of (\ref{TRP2}), (\ref{TPP2}),
(\ref{TPR2}) and (\ref{TRR2}). This can be written as
\begin{eqnarray}
\bm{{\cal T}}&=&\bm{{\cal T}}^{PP}+\bm{{\cal T}}^{RP}+\bm{{\cal
T}}^{PR}+\bm{{\cal T}}^{RR}\nonumber\\
&=&\bm{{\cal T}}^{PP}-\bm{T}_R\bm{G}^0_{BS}\bm{{B}}\bm{{\cal
T}}^{PP}-\bm{{\cal
T}}^{PP}\bm{G}^0_{BS}\bm{{B}}\bm{T}_R+\bm{T}_R+\bm{T}_R\bm{G}^0_{BS}\bm{{B}}
\bm{{\cal T}}^{PP}\bm{G}^0_{BS}\bm{{B}}\bm{T}_R\nonumber\\
&=&\bm{T}_R+\left(\bm{1}-\bm{T}_R\bm{G}^0_{BS}\bm{{B}}\right)
\bm{{\cal
T}}^{PP}\left(\bm{1}-\bm{G}^0_{BS}\bm{{B}}\bm{T}_R\right)\,.\label{Ttotal}
\end{eqnarray}

\section{Wave Functions}

We also need a similar separation for the scattering state of
particle 1 with the bound state of particles 2 and 3 and for the
three-body bound state. To obtain the former, consider the
left-handed propagator defined by (\ref{GL_BS}). Using
(\ref{Ttotal}), this can be rewritten as
\begin{eqnarray}
\bm{{\cal G}}_L&=&\bm{{G}}^0_{BS}-\bm{{G}}^0_{BS}
\left(\bm{{1+B}}\right)\bm{{\cal T}}\bm{{G}}^0_{BS}\nonumber\\
&=&\bm{{G}}^0_{BS}-\bm{{G}}^0_{BS}
\left(\bm{{1+B}}\right)\bm{T}_R\bm{{G}}^0_{BS}\nonumber\\
&&\quad-\bm{{G}}^0_{BS}
\left(\bm{{1+B}}\right)\left(\bm{1}-\bm{T}_R\bm{G}^0_{BS}\bm{{B}}\right)
\bm{{\cal
T}}^{PP}\left(\bm{1}-\bm{G}^0_{BS}\bm{{B}}\bm{T}_R\right)\bm{{G}}^0_{BS}\nonumber\\
&=&\bm{{G}}^0_{BS}-\bm{{G}}^0_{BS}
\left(\bm{{1+B}}\right)\bm{T}_R\bm{{G}}^0_{BS}\nonumber\\
&&\quad-\bm{{G}}^0_{BS}
\left(\bm{{1+B}}\right)\left(\bm{1}-\bm{T}_R\bm{G}^0_{BS}\bm{{B}}\right)
iG_1^{-1}\bm{D}^1\left|\Gamma^{(2)1}\right>i{\cal Q}^1(-iG_{OPT})(i{\cal Q}^1)\nonumber\\
&&\qquad\times\left<\Gamma^{(2)1}\right|{\bm{D}^1}^TiG_1^{-1}
\left(\bm{1}-\bm{G}^0_{BS}\bm{{B}}\bm{T}_R\right)\bm{{G}}^0_{BS}\nonumber\\
&=&\bm{{G}}^0_{BS}-\bm{{G}}^0_{BS}
\left(\bm{{1+B}}\right)\bm{T}_R\bm{{G}}^0_{BS}\nonumber\\
&&\quad+i\bm{{G}}^0_{BS}
\left(\bm{{1+B}}\right)\left(\bm{1}-\bm{T}_R\bm{G}^0_{BS}\bm{{B}}\right)
iG_1^{-1}\bm{D}^1\left|\Gamma^{(2)1}\right>\nonumber\\
&&\quad\times\left(G_1 -G_1 V_{OPT}G_{OPT}\right)i{\cal
Q}^1\left<\Phi^{(2)1}\right|{\bm{D}^1}^T
\left(\bm{1}-\bm{{B}}\bm{T}_R\bm{{G}}^0_{BS}\right)\,.
\end{eqnarray}
From the residue of the pole contribution to  $G_1$, we can identify
the scattering state as
\begin{eqnarray}
\left<\bm{\Phi}^{1(-)}\right|&=&\left<\bm{p}_1,\bm{P}^1\right|\left(1-
V_{OPT}G_{OPT}\right)i{\cal Q}^1\left<\Phi^{(2)1}\right|{\bm{D}^1}^T
\left(\bm{1}-\bm{{B}}\bm{T}_R\bm{{G}}^0_{BS}\right)\nonumber\\
&=&\left<\bm{p}_1,\bm{P}^1\right|\left(1- T_{OPT}G_1\right)i{\cal
Q}^1\left<\Phi^{(2)1}\right|{\bm{D}^1}^T
\left(\bm{1}-\bm{{B}}\bm{T}_R\bm{{G}}^0_{BS}\right)\,,\label{finalstate}
\end{eqnarray}
where $P^1$ is the momentum of the bound state of particles 2 and
3.

The three-body bound-state vertex function, defined by
(\ref{Gamma_BS}), can be written as
\begin{equation}
\left|\bm{{\Gamma}}\right>=-\bm{{\cal V}}\bm{{G}}^0_{BS}\left(
\bm{1}+\bm{{B}}\right)\left|\bm{{\Gamma}}\right>=-\bm{{\cal
M}}\bm{{G}}^0_{BS}\bm{{B}}\left|\bm{{\Gamma}}\right>\,.
\end{equation}
Separating the vertex function into contributions where the last
interaction contains either the pole in the residual parts of the
scattering matrix for particles 2 and 3 gives
\begin{eqnarray}
\left|\bm{{\Gamma}}_P\right>&=&-\bm{{\cal
M}}_P\bm{{G}}^0_{BS}\bm{{B}}\left|\bm{{\Gamma}}\right>\label{GammaP1}\\
\left|\bm{{\Gamma}}_R\right>&=&-\bm{{\cal
M}}_R\bm{{G}}^0_{BS}\bm{{B}}\left|\bm{{\Gamma}}\right>\,.\label{GammaR1}
\end{eqnarray}
Using the definition of the pole contribution,
\begin{eqnarray}
\left|\bm{{\Gamma}}_P\right>&=&\bm{D}^1\left|\Gamma^{(2)1}\right>iG_1^{-1}i{\cal
Q}^1
\left<\Gamma^{(2)1}\right|{\bm{D}^1}^T\bm{{G}}^0_{BS}\bm{{B}}\left|\bm{{\Gamma}}\right>\nonumber\\
&=&\bm{D}^1\left|\Gamma^{(2)1}\right>i{\cal Q}^1
\left<\Phi^{(2)1}\right|{\bm{D}^1}^T\bm{{B}}\left|\bm{{\Gamma}}\right>\,.\label{GammaP2}
\end{eqnarray}
The remaining part of the vertex function is
\begin{eqnarray}
\left|\bm{{\Gamma}}_R\right>&=&-\left({\bm{T}}_R+\bm{T}_R\bm{G}^0_{BS}\bm{{B}}\bm{{\cal
M}}_R\right)\bm{{G}}^0_{BS}\bm{{B}}\left|\bm{{\Gamma}}\right>\nonumber\\
&=&-{\bm{T}}_R\bm{{G}}^0_{BS}\bm{{B}}\left|\bm{{\Gamma}}\right>-\bm{T}_R\bm{G}^0_{BS}\bm{{B}}\bm{{\cal
M}}_R\bm{{G}}^0_{BS}\bm{{B}}\left|\bm{{\Gamma}}\right>\nonumber\\
&=&-{\bm{T}}_R\bm{{G}}^0_{BS}\bm{{B}}\left|\bm{{\Gamma}}\right>+\bm{T}_R\bm{G}^0_{BS}\bm{{B}}
\left|\bm{{\Gamma}}_R\right>\nonumber\\
&=&-{\bm{T}}_R\bm{{G}}^0_{BS}\bm{{B}}\left|\bm{{\Gamma}}_P\right>\label{GammaR2}
\end{eqnarray}
Using $\bm{D}^{1T}\bm{B}\bm{D}^1=0$ in (\ref{GammaP2}), substituting
$\left|\bm{{\Gamma}}_R\right>$ from (\ref{GammaR2}) and iterating
once,
\begin{eqnarray}
\left|\bm{{\Gamma}}_P\right>&=&\bm{D}^1\left|\Gamma^{(2)1}\right>i{\cal
Q}^1
\left<\Phi^{(2)1}\right|{\bm{D}^1}^T\bm{{B}}\left|\bm{{\Gamma}}_R\right>\nonumber\\
&=&-\bm{D}^1\left|\Gamma^{(2)1}\right>i{\cal Q}^1
\left<\Phi^{(2)1}\right|{\bm{D}^1}^T\bm{{B}}{\bm{T}}_R\bm{{G}}^0_{BS}\bm{{B}}
\left|\bm{{\Gamma}}_P\right>\nonumber\\
&=&-\bm{D}^1\left|\Gamma^{(2)1}\right>i{\cal Q}^1
\left<\Phi^{(2)1}\right|{\bm{D}^1}^T\bm{{B}}{\bm{T}}_R\bm{{G}}^0_{BS}\bm{{B}}
\bm{D}^1\left|\Gamma^{(2)1}\right>i{\cal Q}^1
\left<\Phi^{(2)1}\right|{\bm{D}^1}^T\bm{{B}}\left|\bm{{\Gamma}}\right>\nonumber\\
&=&-\bm{D}^1\left|\Gamma^{(2)1}\right>i{\cal Q}^1
\left<\Phi^{(2)1}\right|{\bm{D}^1}^T\bm{{B}}{\bm{T}}_R\bm{{B}}
\bm{D}^1\left|\Phi^{(2)1}\right>(-iG_1)i{\cal Q}^1
\left<\Phi^{(2)1}\right|{\bm{D}^1}^T\bm{{B}}\left|\bm{{\Gamma}}\right>\nonumber\\
&=&-i\bm{D}^1\left|\Gamma^{(2)1}\right> V_{OPT}(-iG_1)
\left<\Phi^{(2)1}\right|{\bm{D}^1}^T\bm{{B}}\left|\bm{{\Gamma}}\right>\,.
\end{eqnarray}
Comparing the first and last lines shows that
\begin{equation}
{\cal
Q}^1\left<\Phi^{(2)1}\right|{\bm{D}^1}^T\bm{{B}}\left|\bm{{\Gamma}}\right>=-V_{OPT}(-iG_1)
\left<\Phi^{(2)1}\right|{\bm{D}^1}^T\bm{{B}}\left|\bm{{\Gamma}}\right>\,,\label{GammaP3}
\end{equation}
which is of the form of the bound-state vertex function for the
optical model.

The complete three-body vertex function can now be reconstructed
using (\ref{GammaR2}) and (\ref{GammaP3}) to give
\begin{eqnarray}
\left|\bm{{\Gamma}}\right>&=&\left|\bm{{\Gamma}}_P\right>+\left|\bm{{\Gamma}}_R\right>
=\left(\bm{1}-{\bm{T}}_R\bm{{G}}^0_{BS}\bm{{B}}\right)\left|\bm{{\Gamma}}_P\right>\nonumber\\
&=&\left(\bm{1}-{\bm{T}}_R\bm{{G}}^0_{BS}\bm{{B}}\right)\bm{D}^1\left|\Gamma^{(2)1}\right>i{\cal
Q}^1
\left<\Phi^{(2)1}\right|{\bm{D}^1}^T\bm{{B}}\left|\bm{{\Gamma}}\right>\,.
\end{eqnarray}
The Bethe-Salpeter wave function is then
\begin{equation}
\left|\bm{\Psi}\right>=\bm{G}^0_{BS}\left|\bm{{\Gamma}}\right>
=\left(\bm{1}-\bm{{G}}^0_{BS}{\bm{T}}_R\bm{{B}}\right)\bm{D}^1\left|\Phi^{(2)1}\right>(-iG_1)i{\cal
Q}^1
\left<\Phi^{(2)1}\right|{\bm{D}^1}^T\bm{{B}}\left|\bm{{\Gamma}}\right>
\end{equation}
and
\begin{equation}
\Psi_{OPT}=(-iG_1)i{\cal Q}^1
\left<\Phi^{(2)1}\right|{\bm{D}^1}^T\bm{{B}}\left|\bm{{\Gamma}}\right>
\end{equation}
can be identified as the optical model wave function.

\section{Electromagnetic Current Matrix Element}\label{sec:current}

\begin{figure}
\centerline{\includegraphics[width=4in]{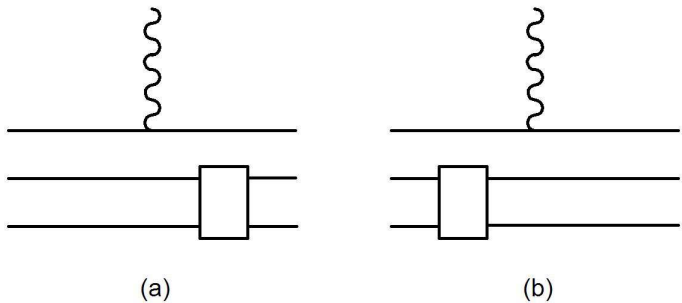}}
\caption{These Feynman diagrams represent contributions to the
seven-point function. The particles are label 1 to 3 from top to
bottom. The rectangles represent two-body kernels.} \label{current1}
\end{figure}

\begin{figure}
\centerline{\includegraphics[width=4in]{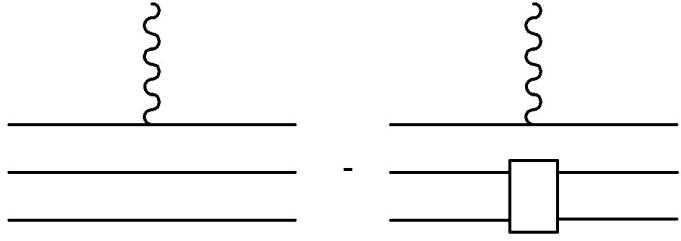}}
\caption{Feynman diagrams representing the correction to the current
operator to correct for double counting.} \label{current2}
\end{figure}

At this point it is necessary to deal with a problem that is occurs
in describing the current matrix element for the Bethe-Salpeter
equation that is not present in the usual nonrelativistic approach.
Some care must be taken with defining the Bethe-Salpeter current
operator if the matrix elements are to be of the form
\begin{equation}
{\cal
J}^\mu=\left<\bm{\Psi}_f\right|\bm{J}^\mu\left|\bm{\Psi}_i\right>\,.
\label{emme1}
\end{equation}
Consider the contribution from the absorption of a virtual photon on
particle 1. The initial state wave function can produce a
contribution described by Fig. \ref{current1}a, while the final
state wave function can produce a contribution described Fig.
\ref{current1}b. Since these are Feynman diagrams and topologically
equivalent, these two diagrams give identical identical
contributions and including both contributions will double count.
Therefore, in order to write the matrix element element in the
symmetric form (\ref{emme1}), it is necessary to correct the current
operator to correct for the double counting.
This can be done by replacing the one-body current by the currents
represented by Fig. \ref{current2}. This was pointed out in
\cite{kb97II,kb99} and is included in the definition of the
Bethe-Salpeter effective current operator defined in \cite{3NCur}
given by (\ref{Jtint}) and (\ref{Jeffmat}).

Now consider the electromagnetic current matrix element for
ejecting particle 1 from the bound state into the continuum state
where particles 2 and 3 remain bound. This is
\begin{eqnarray}
{\cal J}^\mu&=&\left<\bm{\Phi}^{1(-)}\right|\bm{J}^\mu_{\rm
eff}\left|\bm{\Psi}\right>\nonumber\\
&=&\left<\bm{p}_1,\bm{P}^1\right|\left(1-
T_{OPT}(-iG_1)\right)i{\cal Q}^1\left<\Phi^{(2)1}\right|{\bm{D}^1}^T
\left(\bm{1}-\bm{{B}}\bm{T}_R\bm{{G}}^0_{BS}\right)\bm{J}^\mu_{\rm
eff}\nonumber\\
&&\qquad\left(\bm{1}-\bm{{G}}^0_{BS}{\bm{T}}_R\bm{{B}}\right)\bm{D}^1\left|\Phi^{(2)1}\right>(-iG_1)i{\cal
Q}^1
\left<\Phi^{(2)1}\right|{\bm{D}^1}^T\bm{{B}}\left|\bm{{\Gamma}}\right>\nonumber\\
&=&\left<\bm{p}_1,\bm{P}^1\right|\left(1-
T_{OPT}(-iG_1)\right)J^\mu_{OPT}(-iG_1)
\left<\Phi^{(2)1}\right|{\bm{D}^1}^T\bm{{B}}\left|\bm{{\Gamma}}\right>\,,
\end{eqnarray}
where
\begin{equation}
J^\mu_{OPT}=i{\cal Q}^1\left<\Phi^{(2)1}\right|{\bm{D}^1}^T
\left(\bm{1}-\bm{{B}}\bm{T}_R\bm{{G}}^0_{BS}\right)\bm{J}^\mu_{\rm
eff}\left(\bm{1}-\bm{{G}}^0_{BS}{\bm{T}}_R\bm{{B}}\right)\bm{D}^1\left|\Phi^{(2)1}\right>i{\cal
Q}^1\label{Jopt}
\end{equation}
and $\bm{J}^\mu_{\rm eff}$ is defined by (\ref{Jeffmat}).

Considerable care must be taken in evaluating this expression. To
simplify the derivation, we have used the operator $i{\cal Q}^1$ to
place the bound state on shell. Operators of this type were
introduced in \cite{norm} and elaborated in \cite{2NCur} and
\cite{3NCur}. This is a very singular operator and must be treated
with extreme care.  In particular this operator is not associative
and its evaluation depends on its context in the evaluation of
physical quantities. To see this consider the two-body scattering
matrix for particles 2 and 3 given by
\begin{equation}
M^1=V^1-V^1G^1M^1=V^1-M^1G^1V^1\,.
\end{equation}
The second form of this equation can be solved to give
\begin{equation}
V^1=M^1+M^1G^1V^1\,.
\end{equation}
Substituting this into the first form leads to the nonlinear form of
the equation for the scattering matrix
\begin{eqnarray}
M^1&=&V^1-M^1G^1M^1-M^1G^1V^1G^1M^1=V^1-M^1G^1{G^1}^{-1}G^1M^1-M^1G^1V^1G^1M^1\nonumber\\
&=&V^1-M^1G^1({G^1}^{-1}+V^1)G^1M^1=V^1-M^1G^1{{\cal
G}^1}^{-1}G^1M^1\,.
\end{eqnarray}
Note that since ${{\cal G}^1}^{-1}$ must vanish at the bound state
pole, both sides of this equation have simple pole at this point.
The residues of these poles give
\begin{equation}
-\left|\Gamma^{(2)1}\right>i{\cal Q}^1\left<\Gamma^{(2)1}\right|
=-\left|\Gamma^{(2)1}\right>i{\cal
Q}^1\left<\Gamma^{(2)1}\right|G^1{{\cal
G}^1}^{-1}G^1\left|\Gamma^{(2)1}\right>i{\cal
Q}^1\left<\Gamma^{(2)1}\right|\,.
\end{equation}
This requires that
\begin{equation}
i{\cal Q}^1=i{\cal Q}^1\left<\Gamma^{(2)1}\right|G^1{{\cal
G}^1}^{-1}G^1\left|\Gamma^{(2)1}\right>i{\cal Q}^1=i{\cal
Q}^1\left<\Phi^{(2)1}\right|{{\cal
G}^1}^{-1}\left|\Phi^{(2)1}\right>i{\cal Q}^1\,.\label{OpProd}
\end{equation}
Clearly, if we choose to group ${{\cal G}^1}^{-1}$ with either the
first or last occurrence of ${\cal Q}^1$ on the right had side of
this equation, the right-hand side will vanish and the equation will
be violated. This means that the operators on the right-hand side
must be evaluate as a whole without attempting to evaluate them in a
pair-wise manner when appearing in this context.

As a first step in simplifying the optical model current operator,
consider the effective current define in (\ref{Jeffmat}) as
\begin{eqnarray}
\bm{J}^\mu_{\rm eff}&=&J^{(1)\mu}\left(\bm{1}+\bm{B}\right)+
\left(\bm{1}+\bm{B}\right)\bm{J}^\mu_{\rm
int}\left(\bm{1}+\bm{B}\right)=J^{(1)\mu}\bm{D}\bm{D^T}+
\bm{D}\bm{D^T}\bm{J}^\mu_{\rm int}\bm{D}\bm{D^T}\nonumber\\
&=&\bm{D}\left(J^{1\mu}+J^{2\mu}+J^{3\mu}+J^{0\mu}_{\rm
int}+J^{1\mu}_{\rm int}+J^{2\mu}_{\rm int}+J^{3\mu}_{\rm
int}\right)\bm{D^T}\nonumber\\
&=&\bm{D}\left(J^{1\mu}+iV^1J^\mu_1+J^{2\mu}+J^{3\mu}+J^{0\mu}_{\rm
int}+J^{1\mu}_{\rm ex}+J^{2\mu}_{\rm int}+J^{3\mu}_{\rm
int}\right)\bm{D^T}\,,
\end{eqnarray}
where we have used the identity $\bm{1}+\bm{B}=\bm{D}\bm{D}^T$ with
\begin{equation}
\bm{D}=\left(
\begin{array}{c}
1\\1\\1\\1
\end{array}
\right)\, .
\end{equation}

Consider
\begin{equation}
J^{1\mu}+J^{1\mu}_{\rm int}=J^{1\mu}+iJ^\mu_1 V^1+J^1_{ex}\,
\end{equation}
where $J^1_{ex}$ is the two-body current for particles 2 and 3 and
the remaining terms contain only one-body currents. The one-body
currents can be rewritten as
\begin{equation}
J^{1\mu}+iV^1J^\mu_1=iJ^{\mu}_1{G^1}^{-1}+iV^1J^\mu_1=iJ^{\mu}_1{{\cal
G}^1}^{-1}\,,
\end{equation}
Since these pieces contain the inverse of the interacting propagator
for particles 2 and 3, care must be taken when these contributions
are associated with the operator $i{\cal Q}^1$. For this reason we
need deal with  the contributions of these currents to the optical
model current separately. This gives
\begin{eqnarray}
&&i{\cal Q}^1\left<\Phi^{(2)1}\right|{\bm{D}^1}^T
\left(\bm{1}-\bm{{B}}\bm{T}_R\bm{{G}}^0_{BS}\right)\bm{D}iJ^{\mu}_1{{\cal
G}^1}^{-1}\bm{D}^T\left(\bm{1}-\bm{{G}}^0_{BS}{\bm{T}}_R\bm{{B}}\right)\bm{D}^1\left|\Phi^{(2)1}\right>i{\cal
Q}^1\nonumber\\
&&\quad=i{\cal Q}^1\left<\Phi^{(2)1}\right|iJ^{\mu}_1{{\cal
G}^1}^{-1}\left|\Phi^{(2)1}\right>i{\cal Q}^1-i{\cal
Q}^1\left<\Phi^{(2)1}\right|iJ^{\mu}_1{{\cal
G}^1}^{-1}\bm{D}^T\bm{{G}}^0_{BS}{\bm{T}}_R\bm{{B}}{\bm{D}^1}\left|\Phi^{(2)1}\right>i{\cal
Q}^1\nonumber\\
&&\quad-i{\cal
Q}^1\left<\Phi^{(2)1}\right|{\bm{D}^1}^T\bm{{B}}\bm{T}_R\bm{{G}}^0_{BS}\bm{D}iJ^{\mu}_1{{\cal
G}^1}^{-1}\left|\Phi^{(2)1}\right>i{\cal Q}^1\nonumber\\
&&\quad+i{\cal
Q}^1\left<\Phi^{(2)1}\right|{\bm{D}^1}^T\bm{{B}}\bm{T}_R\bm{{G}}^0_{BS}\bm{D}iJ^{\mu}_1{{\cal
G}^1}^{-1}\bm{D}^T\bm{{G}}^0_{BS}{\bm{T}}_R\bm{{B}}{\bm{D}^1}\left|\Phi^{(2)1}\right>i{\cal
Q}^1\nonumber\\
&&\quad=iJ^{\mu}_1i{\cal Q}^1+i{\cal
Q}^1\left<\Phi^{(2)1}\right|{\bm{D}^1}^T\bm{{B}}\bm{T}_R\bm{{G}}^0_{BS}\bm{D}iJ^{\mu}_1{{\cal
G}^1}^{-1}\bm{D}^T\bm{{G}}^0_{BS}{\bm{T}}_R\bm{{B}}{\bm{D}^1}\left|\Phi^{(2)1}\right>i{\cal
Q}^1\,.
\end{eqnarray}
This has been simplified using (\ref{OpProd}) and the identities
\begin{equation}
i{\cal Q}^1{{\cal G}^1}^{-1}{\cal O}={\cal O}{{\cal G}^1}^{-1}i{\cal
Q}^1=0\,,\label{OpProd2}
\end{equation}
where ${\cal O}$ is any operator other than $i{\cal Q}^1$ or ${{\cal
G}^1}^{-1}$. The complete optical model current operator therefore
reduces to
\begin{eqnarray}
J^\mu_{OPT}&=&iJ^{\mu}_1i{\cal Q}^1+i{\cal
Q}^1\left<\Phi^{(2)1}\right|\bigl[
{\bm{D}^1}^T\bm{{B}}\bm{T}_R\bm{{G}}^0_{BS}\bm{D}iJ^{\mu}_1{{\cal
G}^1}^{-1}\bm{D}^T\bm{{G}}^0_{BS}{\bm{T}}_R\bm{{B}}{\bm{D}^1}\nonumber\\
&&+{\bm{D}^1}^T
\left(\bm{1}-\bm{{B}}\bm{T}_R\bm{{G}}^0_{BS}\right)\bm{D}\left(J^{2\mu}+J^{3\mu}+J^{0\mu}_{\rm
int}+J^{1\mu}_{\rm ex}+J^{2\mu}_{\rm int}+J^{3\mu}_{\rm
int}\right)\nonumber\\
&&\times\bm{D}^T\left(\bm{1}
-\bm{{G}}^0_{BS}{\bm{T}}_R\bm{{B}}\right)\bm{D}^1\bigr]\left|\Phi^{(2)1}\right>i{\cal
Q}^1\,.
\end{eqnarray}
Note that the first term yields the usual impulse approximation
while the remaining contribution corresponds to a considerable
number of Feynman diagrams.

To show that the optical model current is conserved we need the
identities
\begin{eqnarray}
q_\mu J^{\mu}_1&=&[e_1(q),G_1^{-1}]\\
q_\mu J^{1\mu}_{\rm ex}&=&[e_2(q)+e_3(q),{\cal V}^1]\\
q_\mu J^{i\mu}&=&[e_i(q),{G^0_{BS}}^{-1}]\\
q_\mu J^{i\mu}_{\rm int}&=&[e_T(q),{\cal V}^i]\,
\end{eqnarray}
where $e_i(q)$ combines the charge operator for particle $i$ with a
four-momentum shift operator and
\begin{equation}
e_T(q)=e_1(q)+e_2(q)+e_3(q)\,.
\end{equation}
Using these identities, contraction of the optical model current
with the four-momentum transfer $q^\mu$ gives
\begin{eqnarray}
q_\mu J^\mu_{OPT}&=&i[e_1(q),G_1^{-1}]i{\cal Q}^1+i{\cal
Q}^1\left<\Phi^{(2)1}\right|\bigl[
{\bm{D}^1}^T\bm{{B}}\bm{T}_R\bm{{G}}^0_{BS}\bm{D}i[e_1(q),G_1^{-1}]{{\cal
G}^1}^{-1}\bm{D}^T\bm{{G}}^0_{BS}{\bm{T}}_R\bm{{B}}{\bm{D}^1}\nonumber\\
&&+{\bm{D}^1}^T
\left(\bm{1}-\bm{{B}}\bm{T}_R\bm{{G}}^0_{BS}\right)\bm{D}\bigl([e_2(q)+e_3(q),{G^0_{BS}}^{-1}]
+[e_2(q)+e_3(q),{\cal V}^1]\nonumber\\
&&+[e_T(q),{\cal V}^0+{\cal V}^2+{\cal
V}^3]\bigr)\bm{D}^T\left(\bm{1}
-\bm{{G}}^0_{BS}{\bm{T}}_R\bm{{B}}\right)\bm{D}^1\bigr]\left|\Phi^{(2)1}\right>i{\cal
Q}^1\nonumber\\
&=&i[e_1(q),G_1^{-1}]i{\cal Q}^1+i{\cal
Q}^1\left<\Phi^{(2)1}\right|\bigl[
{\bm{D}^1}^T\bm{{B}}\bm{T}_R\bm{{G}}^0_{BS}\bm{D}i[e_1(q),G_1^{-1}]{{\cal
G}^1}^{-1}\bm{D}^T\bm{{G}}^0_{BS}{\bm{T}}_R\bm{{B}}{\bm{D}^1}\nonumber\\
&&+{\bm{D}^1}^T
\left(\bm{1}-\bm{{B}}\bm{T}_R\bm{{G}}^0_{BS}\right)\bm{D}\bigl([e_T(q),{G^0_{BS}}^{-1}]
+[e_T(q),{\cal V}^0+{\cal V}^1+{\cal V}^2+{\cal
V}^3]\bigr)\nonumber\\
&&\times\bm{D}^T\left(\bm{1}
-\bm{{G}}^0_{BS}{\bm{T}}_R\bm{{B}}\right)\bm{D}^1\nonumber\\
&&-{\bm{D}^1}^T
\left(\bm{1}-\bm{{B}}\bm{T}_R\bm{{G}}^0_{BS}\right)\bm{D}[e_1(q),{G^0_{BS}}^{-1}+{\cal
V}^1]\bm{D}^T\nonumber\\
&&\times\left(\bm{1}
-\bm{{G}}^0_{BS}{\bm{T}}_R\bm{{B}}\right)\bm{D}^1
\bigr]\left|\Phi^{(2)1}\right>i{\cal Q}^1
\end{eqnarray}
Using
\begin{equation}
iG_1^{-1}{{\cal G}^1}^{-1}={G^0_{BS}}^{-1}+{\cal V}^1\,,
\end{equation}
this can be simplified as
\begin{eqnarray}
q_\mu J^\mu_{OPT}&=&i[e_1(q),G_1^{-1}]i{\cal Q}^1+i{\cal
Q}^1\left<\Phi^{(2)1}\right|\bigl[{\bm{D}^1}^T
\left(\bm{1}-\bm{{B}}\bm{T}_R\bm{{G}}^0_{BS}\right)\bm{D}
\nonumber\\
&&\times\bigl([e_T(q),{G^0_{BS}}^{-1}] +[e_T(q),{\cal V}^0+{\cal
V}^1+{\cal V}^2+{\cal V}^3]\bigr)\bm{D}^T\left(\bm{1}
-\bm{{G}}^0_{BS}{\bm{T}}_R\bm{{B}}\right)\bm{D}^1\nonumber\\
&&-[e_1(q),{G^0_{BS}}^{-1}+{\cal V}^1]
+[e_1(q),{G^0_{BS}}^{-1}+{\cal
V}^1]\bm{D}^T\bm{{G}}^0_{BS}{\bm{T}}_R\bm{{B}}{\bm{D}^1}\nonumber\\
&&+{\bm{D}^1}^T\bm{{B}}\bm{T}_R\bm{{G}}^0_{BS}\bm{D}[e_1(q),{G^0_{BS}}^{-1}+{\cal
V}^1] \bigr]\left|\Phi^{(2)1}\right>i{\cal Q}^1\,.
\end{eqnarray}
From (\ref{OpProd2}),
\begin{equation}
i{\cal Q}^1[e_1(q),{G^0_{BS}}^{-1}+{\cal V}^1]i{\cal Q}^1=0\,.
\end{equation}
So,
\begin{eqnarray}
q_\mu J^\mu_{OPT}&=&i[e_1(q),G_1^{-1}]i{\cal Q}^1+i{\cal
Q}^1\left<\Phi^{(2)1}\right|\bigl[{\bm{D}^1}^T
\left(\bm{1}-\bm{{B}}\bm{T}_R\bm{{G}}^0_{BS}\right)\bm{D}
\nonumber\\
&&\times\bigl([e_T(q),{G^0_{BS}}^{-1}] +[e_T(q),{\cal V}^0+{\cal
V}^1+{\cal V}^2+{\cal V}^3]\bigr)\bm{D}^T\left(\bm{1}
-\bm{{G}}^0_{BS}{\bm{T}}_R\bm{{B}}\right)\bm{D}^1\nonumber\\
&& +e_1(q)\left({G^0_{BS}}^{-1}+{\cal
V}^1\right)\bm{D}^T\bm{{G}}^0_{BS}{\bm{T}}_R\bm{{B}}{\bm{D}^1}\nonumber\\
&&-{\bm{D}^1}^T\bm{{B}}\bm{T}_R\bm{{G}}^0_{BS}\bm{D}\left({G^0_{BS}}^{-1}+{\cal
V}^1\right)e_1(q) \bigr]\left|\Phi^{(2)1}\right>i{\cal Q}^1\,.
\end{eqnarray}

No consider
\begin{equation}
\left({G^0_{BS}}^{-1}+{\cal
V}^1\right)\bm{D}^T\bm{{G}}^0_{BS}{\bm{T}}_R\bm{{B}}{\bm{D}^1}
={\bm{D}^1}^T[(\bm{1}+\bm{B}){\bm{T}}_R\bm{{B}}+\bm{{\cal
V}}(\bm{1}+\bm{B})\bm{{G}}^0_{BS}{\bm{T}}_R\bm{{B}}]{\bm{D}^1}\,.
\end{equation}
This can be simplified using
\begin{eqnarray}
\bm{{\cal V}}\bm{{G}}^0_{BS}{\bm{T}}_R\bm{{B}}&=&\bm{{\cal
V}}\bm{{G}}^0_{BS}\bm{{\cal
M}}_R\left(\bm{1}-\bm{{G}}^0_{BS}\bm{B}\bm{T}_R\right)\bm{{B}}\nonumber\\
&=&\bm{{\cal V}}\bm{{G}}^0_{BS}\left(\bm{{\cal M}}-\bm{{\cal
M}}_P\right)\left(\bm{1}-\bm{{G}}^0_{BS}\bm{B}\bm{T}_R\right)\nonumber\\
&=&\left(\bm{{\cal V}}-\bm{{\cal M}}+\bm{{\cal
M}}_P\right)\left(\bm{1}-\bm{{G}}^0_{BS}\bm{B}\bm{T}_R\right)\bm{{B}}\nonumber\\
&=&\left(\bm{{\cal V}}-\bm{{\cal M}}_R\right)
\left(\bm{1}-\bm{{G}}^0_{BS}\bm{B}\bm{T}_R\right)\bm{{B}}\nonumber\\
&=&\bm{{\cal V}}\bm{{B}}-\bm{{\cal
V}}\bm{{G}}^0_{BS}\bm{B}\bm{T}_R\bm{B}-\bm{T}_R\bm{B}
\end{eqnarray}
such that
\begin{eqnarray}
\left({G^0_{BS}}^{-1}+{\cal
V}^1\right)\bm{D}^T\bm{{G}}^0_{BS}{\bm{T}}_R\bm{{B}}{\bm{D}^1}
&=&{\bm{D}^1}^T[(\bm{1}+\bm{B}){\bm{T}}_R\bm{{B}}+ \bm{{\cal
V}}\bm{{B}}-\bm{{\cal
V}}\bm{{G}}^0_{BS}\bm{B}\bm{T}_R\bm{B}-\bm{T}_R\bm{B}\nonumber\\
&&+ \bm{{\cal
V}}\bm{B}\bm{{G}}^0_{BS}{\bm{T}}_R\bm{{B}}]{\bm{D}^1}\nonumber\\
&=&{\bm{D}^1}^T[\bm{{B}}{\bm{T}}_R\bm{{B}}+ \bm{{\cal
V}}\bm{{B}}]{\bm{D}^1}={\bm{D}^1}^T\bm{{B}}{\bm{T}}_R\bm{{B}}{\bm{D}^1}\,.
\end{eqnarray}
Similarly,
\begin{equation}
{\bm{D}^1}^T\bm{{B}}{\bm{T}}_R\bm{{G}}^0_{BS}\bm{D}\left({G^0_{BS}}^{-1}+{\cal
V}^1\right) ={\bm{D}^1}^T\bm{{B}}{\bm{T}}_R\bm{{B}}{\bm{D}^1}\,.
\end{equation}

Note also that
\begin{eqnarray}
\bm{D}[e_T(q),{G^0_{BS}}^{-1}+{\cal V}^0+{\cal V}^1+{\cal V}^2+{\cal
V}^3]\bm{D}^T&=&e_T(q)(\bm{1}+\bm{B})[{\bm{G}^0_{BS}}^{-1}+\bm{{\cal
V}}(\bm{1}+\bm{B})]\nonumber\\
&&-[{\bm{G}^0_{BS}}^{-1}+(\bm{1}+\bm{B})\bm{{\cal
V}}](\bm{1}+\bm{B}) e_T(q)\,,
\end{eqnarray}
and
\begin{eqnarray}
&&[{\bm{G}^0_{BS}}^{-1}+\bm{{\cal V}}(\bm{1}+\bm{B})]\left(\bm{1}
-\bm{{G}}^0_{BS}{\bm{T}}_R\bm{{B}}\right)\bm{D}^1\nonumber\\
&&=[{\bm{G}^0_{BS}}^{-1}+\bm{{\cal
V}}(\bm{1}+\bm{B})-{\bm{T}}_R\bm{{B}}-\bm{{\cal
V}}(\bm{1}+\bm{B})\bm{{G}}^0_{BS}{\bm{T}}_R\bm{{B}}]\bm{D}^1\nonumber\\
&&=[{\bm{G}^0_{BS}}^{-1}+\bm{{\cal
V}}(\bm{1}+\bm{B})-{\bm{T}}_R\bm{{B}}-\bm{{\cal
V}}\bm{{B}}+\bm{{\cal
V}}\bm{{G}}^0_{BS}\bm{B}\bm{T}_R\bm{B}+\bm{T}_R\bm{B}-\bm{{\cal
V}}\bm{B})\bm{{G}}^0_{BS}{\bm{T}}_R\bm{{B}}]\bm{D}^1\nonumber\\
&&=[{\bm{G}^0_{BS}}^{-1}+\bm{{\cal
V}}]\bm{D}^1=\bm{D}^1[{{G}^0_{BS}}^{-1}+{{\cal V}^1}]\,.
\end{eqnarray}
Similarly,
\begin{equation}
{\bm{D}^1}^T
\left(\bm{1}-\bm{{B}}\bm{T}_R\bm{{G}}^0_{BS}\right)[{\bm{G}^0_{BS}}^{-1}+(\bm{1}+\bm{B})\bm{{\cal
V}}]=[{{G}^0_{BS}}^{-1}+{{\cal V}^1}]{\bm{D}^1}^T\,.
\end{equation}

Using these identities we can rewrite
\begin{eqnarray}
q_\mu J^\mu_{OPT}&=&i[e_1(q),G_1^{-1}]i{\cal Q}^1\nonumber\\
&&+i{\cal Q}^1\left<\Phi^{(2)1}\right|\bigl[{\bm{D}^1}^T
\left(\bm{1}-\bm{{B}}\bm{T}_R\bm{{G}}^0_{BS}\right)e_T(q)(\bm{1}+\bm{B})
\bm{D}^1[{{G}^0_{BS}}^{-1}+{{\cal V}^1}]\nonumber\\
&&-[{{G}^0_{BS}}^{-1}+{{\cal V}^1}]{\bm{D}^1}^T(\bm{1}+\bm{B})
e_T(q)\left(\bm{1}
-\bm{{G}}^0_{BS}{\bm{T}}_R\bm{{B}}\right)\bm{D}^1\nonumber\\
&& +e_1(q){\bm{D}^1}^T\bm{{B}}{\bm{T}}_R\bm{{B}}{\bm{D}^1}
-{\bm{D}^1}^T\bm{{B}}{\bm{T}}_R\bm{{B}}{\bm{D}^1}e_1(q)
\bigr]\left|\Phi^{(2)1}\right>i{\cal Q}^1\nonumber\\
&=&i[e_1(q),G_1^{-1}]i{\cal Q}^1+e_1(q)i{\cal
Q}^1\left<\Phi^{(2)1}\right|{\bm{D}^1}^T\bm{{B}}{\bm{T}}_R\bm{{B}}{\bm{D}^1}\left|\Phi^{(2)1}\right>i{\cal
Q}^1\nonumber\\
&&-i{\cal
Q}^1\left<\Phi^{(2)1}\right|{\bm{D}^1}^T\bm{{B}}{\bm{T}}_R\bm{{B}}{\bm{D}^1}\left|\Phi^{(2)1}\right>i{\cal
Q}^1e_1(q)\nonumber\\
&=&i[e_1(q),G_1^{-1}]i{\cal Q}^1+[e_1(q),iV_{OPT}]
\end{eqnarray}
Or,
\begin{equation}
q_\mu J^\mu_{OPT}=i[e_1(q),G_{OPT}^{-1}]\,.
\end{equation}
This is the Ward-Takahashi identity for the optical model current
and along with the wave equations for the optical model wave
functions guaranties that the current matrix elements will be
conserved.

\section{Conclusions}

We have shown that the three-body Bethe-Salpeter equation can be
reduced to an effective two-body optical model.  An effective
current appropriate to this model has been constructed. This current
is shown to satisfy a Ward-Takahashi identity involving the optical
potential which results in conserved current matrix elements. This
conserved current contains a substantial number of contributions not
included in current RDWIA calculations and the contributions of
these extra terms in various kinematical regions need to be
considered carefully.

Although this paper deals with a simple three-body system, extension
of this approach two include additional constituents is possible as
will be described in a subsequent paper.  It may also be useful to
consider limiting cases of this approach to understand its
relationship to the mean field approaches used for most
calculations.

%____________________________________________________________________
\appendix

\section{Review of the Matrix form of the BS equation for Distinguishable Particles}

This appendix contains a short summary of the matrix form of the
three-body Bethe-Salpeter equations and effective current as defined
in \cite{3NCur}.

The three-body Bethe-Salpeter equation can be obtained by examining
the sum of all Feynman diagrams contributing to the three-body
scattering matrix. Contributions to these diagrams can be classified
according to whether the contribution can be separated by cutting
only the three propagators associated with the external legs of the
scattering matrix.  Those diagrams which can not be separated in
this way are three-body irreducible diagrams. The irreducible
diagrams fall into two classes: those where only two of the three
particles are interacting and those where all three particles are
interacting. The sum of all three-body irreducible diagrams is
represented by the kernel $V^0$ and the two-body irreducible
diagrams contribute to the two-body kernels $V^i$  where only
particles $j$ and $k$ (with $j\neq k\neq i$) are interacting.

The complete scattering amplitude can then be written in terms of an
integral equation with the above mentioned kernels.  It is
convenient to express the complete scattering matrix in terms of
subamplitudes $T^{ij}$ where the indices $i$ and $j$ characterize
the subamplitudes according to the character of the last and first
interactions; that is, for $i\neq 0$ ($j\neq 0$), the particles $i$
($j$) are not taking part in the last (first) interaction, $i=0$
($j=0$) means that the last (first) interaction is genuine
three-particle interaction.
%The subamplitudes ${\cal T}^{ij}$ satisfy the following integral
equations (see also ref.\ \cite{norm}):
\begin{equation}
{\cal T}^{ij}= {\cal V}^i\delta_{ij}- {\cal V}^i \bm{G}^0_{BS}
 \sum_{k=0}^3  {\cal T}^{kj}\,,
\end{equation}
where
\begin{equation}
{\cal V}^i=\left\{
\begin{array}{ll}
V^0 & {\rm for}\ i=0\\
V^i iG^{-1}_i & {\rm for}\ i=1,2,3
\end{array}\right.
\end{equation}
and $G^0_{BS}=-G_1G_2G_3$. The form of these equations suggests that
it is convenient to represent this set of equations in a matrix
form. Defining the matrices $(\bm{{\cal V}})_{ij}={\cal V}^i
\delta_{ij}$, $(\bm{{\cal T}})_{ij}= {\cal T}^{ij}$ and
$(\bm{B})_{ik}=1-\delta_{ik}$ for $ i,j= 0,1,2,3$, the three-body
scattering equations can be written as
\begin{equation}
\bm{{\cal T}}
 = \bm{{\cal V}}-\bm{{\cal V}}\bm{G}^0_{BS}(\bm{{1+B}}) \bm{{\cal
T}}=\bm{{\cal V}}-\bm{{\cal T}}(\bm{{1+B}})\bm{G}^0_{BS}\bm{{\cal
V}}
 \, ,
\label{t_BS2}
\end{equation}
where $\bm{{G}}^0_{BS}=G^0_{BS}\bm{1}$.

Numerical solution of these integral equations requires that they
must be put in a form where the kernels are connected or can be made
to be connect by iteration. This is done by reexpressing the
equations in terms of two- and three-body  t-matrices defined in
terms of the corresponding interaction kernels as:
\begin{equation}
\bm{{\cal M}}=\bm{{\cal V}}-\bm{{\cal V}}\bm{G}^0_{BS}\bm{{\cal M}}=
\bm{{\cal V}}-\bm{{\cal M}}\bm{G}^0_{BS}\bm{{\cal V}} \, .
\label{m-matrix}
\end{equation}
The complete t-matrix can then be written as
\begin{equation}
\bm{{\cal T}}= \bm{{\cal M}}-\bm{{\cal M}}\bm{G}^0_{BS}\bm{{B}}
\bm{{\cal T}}=\bm{{\cal M}}-\bm{{\cal T}}\bm{{B}}\bm{G}^0_{BS}
\bm{{\cal M}} \label{t_BS}\,.
\end{equation}

In matrix form, it is necessary to define right- and left-handed
propagators
\begin{equation}
\bm{{\cal G}}_R=\bm{{G}}^0_{BS}-\bm{{G}}^0_{BS} \bm{{\cal T}}
\left(\bm{{1+B}}\right)\bm{{G}}^0_{BS}=\bm{{G}}^0_{BS}-\bm{{G}}^0_{BS}
\bm{{\cal V}}\left(\bm{{1+B}}\right)\bm{{\cal G}}_R \label{GR_BS}
\end{equation}
and
\begin{equation}
\bm{{\cal G}}_L=\bm{{G}}^0_{BS}-\bm{{G}}^0_{BS}
\left(\bm{{1+B}}\right)\bm{{\cal
T}}\bm{{G}}^0_{BS}=\bm{{G}}^0_{BS}-\bm{{\cal G}}_L
\left(\bm{{1+B}}\right)\bm{{\cal V}}\bm{{G}}^0_{BS} \label{GL_BS}\,.
\end{equation}
The inverses of these propagators are
\begin{eqnarray}
\bm{{\cal G}}^{-1}_R&=&({\bm{{G}}^0_{BS}})^{-1}+ \bm{{\cal V}}
\left(\bm{{1+B}}\right)\, ,
\label{GR_BSinv}\\
\bm{{\cal G}}^{-1}_L&=&({\bm{{G}}^0_{BS}})^{-1}+
\left(\bm{{1+B}}\right)\bm{{\cal V}} \label{GL_BSinv}\, .
\end{eqnarray}

The bound state can be obtained from consideration of the
singularities of the t-matrix. and satisfies the equations
\begin{eqnarray}
\left|\bm{{\Gamma}}\right>&=&-\bm{{\cal V}}\bm{{G}}^0_{BS}\left(
\bm{1}+\bm{{B}}\right)\left|\bm{{\Gamma}}\right> \, ,\label{Gamma_BS}\\
\left<\bm{{\Gamma}}\right|&=&-\left<\bm{{\Gamma}}\right|\left(
\bm{1}+\bm{{B}}\right)\bm{{G}}^0_{BS}\bm{{\cal V}} \, .
\end{eqnarray}
Defining the Bethe-Salpeter wave function as
$\left|\bm{{\Psi}}\right>=
\bm{{G}}^0_{BS}\left|\bm{{\Gamma}}\right>$ these can be rewritten as
\begin{eqnarray}
\left[ ({\bm{{G}}^0_{BS}})^{-1}+\bm{{\cal V}}\left(
\bm{1}+\bm{{B}}\right)\right]\left|\bm{{\Psi}}\right>&=&
\bm{{\cal G}}^{-1}_R\left|\bm{{\Psi}}\right>=0 \, , \\
\left<\bm{{\Psi}}\right|\left[ ({\bm{{G}}^0_{BS}})^{-1}+\left(
\bm{1}+\bm{{B}}\right)\bm{{\cal V}}\right]&=&
\left<\bm{{\Psi}}\right|\bm{{\cal G}}^{-1}_L=0 \, .
\end{eqnarray}
The scattering wave functions also satisfy the same equations.

The three-body Bethe-Salpeter current can be determined by
considering all diagrams contributing to the seven-point function
with six legs corresponding to the three incoming and outgoing
particles and one photon leg. By separating the diagrams into
three-particle reducible and irreducible contributions the current
operator can be identified as the sum of all irreducible seven-point
functions. There will be three types of contributions: one-body
contributions where the photon attaches to only one of the
interacting particles, two-body contributions where the photon
attaches internally to a two-body interaction, and three-body
contributions where the photon attaches internally to a three-body
interaction.

The one-body currents are of the form
\begin{equation}
J^{t\mu}= -J^\mu_t G_r^{-1}G_s^{-1} \, , \label{IAcurr}
\end{equation}
where $t=1,2,3$, $r\neq s\neq t$ and $J^\mu_t$ is the vertex for
attaching a photon to particle $t$, for which $q_\mu
J^\mu_t=\left[e_t(q), G_t^{-1}\, \right]$. Each of these currents
satisfies the Ward-Takahashi identity
\begin{equation}
q_\mu J^{t\mu}=\left[e_t(q), (G^0_{BS})^{-1}\, \right] \, .
\end{equation}

The two- and three-body currents follow from attaching a photon line
to all particle lines and into momentum-dependent vertices internal
to the two- and three-body Bethe-Salpeter kernels. Defining
$J^{(2)\mu}_{rs}$ as the two-body current associated with the
two-body contributions to the kernel and $J^{(3)\mu}$ as the
completely connected three-body current, one can write the exchange
current for the three-body system  as
\begin{equation}
J^{t\mu}_{\rm ex}= \left\{ \begin{array}{ll}
J^{(2)\mu}_{rs} iG^{-1}_t, &t=1,2,3 \ \ (r\neq s\neq t) \\
J^{(3)\mu},           &t=0
\end{array}\right.
\end{equation}
which satisfies the Ward-Takahashi identities
\begin{equation}
q_\mu J^{t\mu}_{\rm ex}=\left\{ \begin{array}{ll}
\left[e_r(q)+ e_s(q), {\cal V}^t\right], &t=1,2,3\ \ (r\neq s\neq t)\\
\left[e_T(q), {\cal V}^0\right],         &t=0
\end{array}\right.
\end{equation}
where $e_T(q)= e_1(q)+ e_2(q)+ e_3(q)$.

Following the argument in Section \ref{sec:current}, it is necessary
to include a contribution to the effective current to compensate for
the double counting inherent in the symmetric expression for the
current matrix element. This can be done by defining the interaction
current
\begin{eqnarray}
J_{\rm int}^{t\mu}&=& J^{t\mu}_{ex}+ iV^t J^\mu_t \, \quad
\mbox{for}\ t\neq0 \, ,
\label{Jtint}\\
J_{\rm int}^{0\mu}&=&J_{\rm ex}^{0\mu}= J^{(3)\mu} \, .\label{J0int}
\end{eqnarray}

The matrix form of the effective current is obtained by first
defining the total one-body current as
\begin{equation}
J^{(1)\mu}=\sum_{i=1}^3 J^{i\mu} \, ,
\end{equation}
and defining a diagonal matrix with components defined by
(\ref{Jtint}) and (\ref{J0int}):
\begin{eqnarray}
\bm{J}^\mu_{\rm int}&=& diag\left( J^{0\mu}_{\rm int},
J^{1\mu}_{\rm int},
J^{2\mu}_{\rm int}, J^{3\mu}_{\rm int} \right) \, , \\
q_\mu \bm{J}^\mu_{\rm int} &=&  \left[ e_T(q), \bm{{\cal V}}
\right] \, .
\end{eqnarray}
The effective current can then be identified as
\begin{equation}
\bm{J}^\mu_{\rm eff}=J^{(1)\mu}\left(\bm{{1+B}}\right)+
\left(\bm{{1+B}}\right)\bm{J}^\mu_{\rm int}\left(\bm{{1+B}}\right)
\, . \label{Jeffmat}
\end{equation}
Contraction of the four-momentum transfer with the effective current
gives
\begin{eqnarray}
q_\mu\bm{J}^\mu_{\rm eff}
&=&\left[e_T(q),{\bm{G}^0_{BS}}^{-1}\right]\left(\bm{{1+B}}\right)+
\left(\bm{{1+B}}\right)\left[e_T(q),\bm{{\cal V}}\right]
\left(\bm{{1+B}}\right)\nonumber\\
&=&e_T(q)\left(\bm{{1+B}}\right)\bm{{\cal G}}_R^{-1}-\bm{{\cal
G}}_L^{-1}\left(\bm{{1+B}}\right)e_T(q)\,. \label{WTmat}
\end{eqnarray}
So, using the wave equations, the current will be conserved.


\begin{thebibliography}{99}

\bibitem{PVOW85} A. Picklesimer, J. W. Van~Orden, and S. J.
Wallace, Phys. Rev. C{\bf 32}, 1312 (1985).

\bibitem{PVO87}  A. Picklesimer and J. W. Van~Orden, Phys. Rev.
C{\bf 35}, 266 (1987).

\bibitem{PVO89} A. Picklesimer and J. W. Van~Orden, Phys.\ Rev.\
C{\bf 40}, 290 (1989).

\bibitem{CPVO89a} C. R. Chinn, A. Picklesimer, and J.
W. Van~Orden, Phys.\ Rev.\ C{\bf 40}, 790 (1989).

\bibitem{CPVO89b} C. R. Chinn, A.
Picklesimer and J. W. Van~Orden, Phys.\ Rev.\ C{\bf 40}, 1159
(1989).

\bibitem{CP92}  C.R. Chinn , A. Picklesimer, Nuovo Cim. A105, 1149
(1992).


\bibitem{JOW92} Y. Jin, D.S. Onley, and L.E. Wright, Phys. Rev. C {\bf 45}, 1311
(1992). %(e,e'p)

\bibitem{JO} Y. Jin and D.S. Onley, Phys. Rev. C {\bf 50}, 377 (1994). %(e,e')


\bibitem{KW97} K. S. Kim and L. E. Wright, Phys. Rev. C {\bf 56}, 302 (1997). %(e,e'p) Coulomb effects


\bibitem{KW99} K. S. Kim and L. E. Wright, Phys. Rev. C {\bf 60}, 067604 (1999). %(e,e'p) Coulomb effects


\bibitem{KW03} K. S. Kim and L. E. Wright, Phys. Rev. C {\bf 68}, 027601 (2003). %(e,e'p) and (e,e') medium modifications


\bibitem{USMGC93} J.M. Udias, P. Sarriguren, E. Moya de Guerra, E. Garrido, and
J.A. Caballero, Phys. Rev. C {\bf 48}, 2731 (1993). %(e,e'p) coulomb varied current operators

\bibitem{USMGC95} J.M. Udias, P. Sarriguren, E. Moya de Guerra, E. Garrido, and
J.A. Caballero, Phys. Rev. C {\bf 51}, 3246 (1995). %(e,e'p)

\bibitem{USMGC96} J.M. Udias, P. Sarriguren, E. Moya de Guerra, and J.A. Caballero,
 Phys. Rev. C  {\bf 53}, R1488 (1996). %(e,e'p)


\bibitem{UCMAD99} J. M. Udias, J. A. Caballero, E. Moya de Guerra, J.
E. Amaro, and T. W. Donnelly, Phys. Rev. Lett. {\bf 83}, 5451
(1999). %(e,e'p)

\bibitem{UV00} J. M. Udias and J. R. Vignote, Phys. Rev. C{\bf 62}, 034302
(2000). %(e,e'p) various relativistic currents

\bibitem{UCMVE01} J. M. Udias, J. A. Caballero, E. Moya de Guerra, Javier R.
Vignote, and A. Escuderos, Phys. Rev. C {\bf 64}, 024614 (2001). %(e,e'p)

\bibitem{MCDMU04} M. C. Martinez, J. B. Vignote, J. A. Caballero, T.
W. Donnelly, E. Moya de Guerra and J. M. Udias, Phys. Rev. C{\bf
69}, 034604 (2004). %(e,e'p)

\bibitem{VMCMU04} Javier R. Vignote, M. C. Martínez, J. A. Caballero, E. Moya de
Guerra, and J. M. Udias, Phys. Rev. C {\bf 70}, 044608 (2004). %(e,e'p)


\bibitem{MLJRVU06} M. C. Martinez, P. Lava, N. Jachowicz, J. Ryckebusch, and K.
Vantournhout and J. M. Udias, Phys. Rev. C {\bf 73}, 024607 (2006) %neutrino scattering DWIA and Glauber



\bibitem{MGP01a} A. Meucci, C. Giusti, and F.D. Pacati, Phys. Rev. C {\bf 64}, 014604
(2001). %(e,e'p)

\bibitem{MGP01b} A. Meucci, C. Giusti, and F.D. Pacati, Phys. Rev. C {\bf 64}, 064615
(2001).  %(gamma,N)

\bibitem{MGP02} A. Meucci, C. Giusti and F. D. Pacati,
Phys. Rev. C {\bf 66}, 034610 (2002). %RDWIA and exchange currents

\bibitem{MCGP03} A. Meucci, F. Capuzzi, C. Giusti, and F. D.
Pacati, Phys. Rev. C {\bf 6}7, 054601 (2003). %Inclusive

%
\bibitem{Fissum} K. G. Fissum, {\it et al.}, Phys. Rev. C{\bf 70},
034606 (2004).
%
\bibitem{Dieterich} S. Dieterich, {\it et al.}, Phys. Lett. B {\bf
500}, 47 (2001).
%
\bibitem{Strauch} S. Strach, {\it et al.}, Phys. Rev. Lett. {\bf
91}, 052301 (2003).
%
\bibitem{Aniol} K. A. Aniol, {\it et al.}, Eur. Phys. J. A {\bf 22},
449 (2004).


\bibitem{SBKMV05} R. Schiavilla, O. Benhar, A. Kievshy, L. E.
Marcucci and M. Viviani, Phys. Rev. Lett. {\bf 94}, 072303 (2005).


%
\bibitem{AC79} L. G. Arnold and B. C. Clark, Phys. Lett. {bf 84B},
46 (1979).

\bibitem{ACM79} L. G. Arnold, B. C. Clark and R. L. Mercer, Phys.
Rev. C{\bf 19}, 917 (1979).

\bibitem{ACMS81} L. G. Arnold, B. C. Clark, R. L. Mercer and P. Schwandt, Phys.
Rev. C{\bf 23}, 917 (1981).

\bibitem{SMW93} J. M. Shepard, J. A. McNeil and S. J. Wallace, Phys
Rev. Lett {\bf 50}, 1443 (1983).

\bibitem{CHMRS} B. C. Clark, S. Hama, R. L. Mercer, L. Ray and B. D.
Serot, Phys. Rev. Lett. {\bf 50}, 1644 (1983).

\bibitem{HPTT84} M. V. Hynes, A. Picklesimer, P. C. Tandy and R. M.
Thaler, Phys. Rev. Lett. {\bf 52}, 978 (1984).

\bibitem{BVO90} P.~M. Boucher and J. W.
Van~Orden, Phys. Rev. C{\bf 43}, 582 (1990).

\bibitem{Taylor66} J. G. Taylor, Phys. Rev. {\bf 105}, 1321 (1966).
%
\bibitem{norm} J.~Adam, Jr., F.~Gross, C.~Savkli, and J.~W.~Van Orden,
    Phys.~Rev.~C {\bf 56} (1997) 641.

%
\bibitem{kb97II} A.~N.~Kvinikhidze and B.~Blankleider,
               Phys.~Rev.~C {\bf 56} (1997) 2973.
%
\bibitem{kb99} A.~N.~Kvinikhidze and B.~Blankleider,
               Phys.~Rev.~C {\bf 60} (1999) 044003;
               (1999) 044004.


%
\bibitem{3NCur} J.~Adam, Jr. and J.~W.~Van Orden, Phys. Rev. C{\bf
71}, 034003 (2005).

\bibitem{2NCur} J. Adam, J. W. Van Orden and F. Gross, Nucl. Phys. {\bf A640}, 391
(1998).




\end{thebibliography}
\end{document}